\documentclass[twocolumn,showpacs,preprintnumbers,amsmath,amssymb]{revtex4}

\newcommand{\bib}{\bibitem}
\newcommand{\bea}{\begin{eqnarray}}
\newcommand{\eea}{\end{eqnarray}}
\newcommand{\beq}{\begin{equation}}
\newcommand{\eeq}{\end{equation}}
\newcommand{\non}{\nonumber}

\newcommand{\ep}{\epsilon}

\newcommand{\si}{\sigma}
\newcommand{\Si}{\Sigma}

\newcommand{\up}{\uparrow}
\newcommand{\dn}{\downarrow}

\usepackage{graphicx,epsfig} 
\usepackage{dcolumn} 
\usepackage{bm} 

\begin{document}
\title{Spin-1 Kitaev model in one dimension} 
\author{Diptiman Sen$^1$, R. Shankar$^2$, Deepak Dhar$^3$ and Kabir Ramola$^3$}
\affiliation{$^1$Center for High Energy Physics, Indian Institute of Science, 
Bangalore 560 012, India \\
$^2$The Institute of Mathematical Sciences, CIT Campus, Chennai 600 113, 
India \\
$^3$Department of Theoretical Physics, Tata Institute of Fundamental Research,
Mumbai 400 005, India}
\date{\today}

\begin{abstract} 
We study a one-dimensional version of the Kitaev model on a ring of size $N$, 
in which there is a spin $S > 1/2$ on each site and the Hamiltonian is 
$J \sum_n S^x_n S^y_{n+1}$. The cases where $S$ is integer and 
half-odd-integer are qualitatively different. We show that there is a 
$\mathbb{Z}_2$ valued conserved quantity $W_n$ for each bond $(n,n+1)$
of the system. For integer $S$, the Hilbert space can be decomposed into 
$2^N$ sectors, of unequal sizes. The number of states in most of the sectors 
grows as $d^N$, where $d$ depends on the sector. The largest sector contains 
the ground state, and for this sector, for $S=1$, $d =(\sqrt{5}+1)/2$. We 
carry out exact diagonalization for small systems. The extrapolation of our 
results to large $N$ indicates that the energy gap remains finite in this 
limit. In the ground state sector, the system can be mapped to a spin-1/2 
model. We develop variational wave functions to study the lowest energy states 
in the ground state and other sectors. The first excited state of the system 
is the lowest energy state of a different sector and we estimate its 
excitation energy. We consider a more general Hamiltonian, adding a term 
$\lambda \sum_n W_n$, and show that this has gapless excitations in the range 
$\lambda^c_1 \leq \lambda \leq \lambda^c_2$. We use the variational 
wave functions to study how the ground state energy and the defect density 
vary near the two critical points $\lambda^c_1$ and $\lambda^c_2$.
\end{abstract}

\maketitle

\section{Introduction} 
\label{intro}

In recent years, there have been many studies of quantum spin
systems which are characterized by a high degree of frustration and 
topological order. The word `frustration' here refers to systems with 
competing interactions having a large number of states with energy near the 
minimum energy. Topological order implies the existence of invariants which, 
for topological reasons, are robust against a large class of perturbations. 
Such systems are often associated with a novel structure of the ground
state and low-lying excitations, and are interesting from the point of view 
of possible applications in quantum computation
\cite{kitaev1,levin,kitaev2,sarma,nayak}. A particularly interesting model in
this context is the two-dimensional frustrated spin-1/2 model introduced by
Kitaev \cite{kitaev2}. This model has several fascinating properties which 
have been studied in great detail
\cite{feng,baskaran1,lee,nussinov1,nussinov2,vidal,sengupta}. For instance, the
model and its variants constitute the only known class of spin models in two
dimensions or more dimensions that is fully integrable, being reducible to a 
system of non-interacting Majorana fermions. A similar model, called the 
compass model, although not exactly solvable, was introduced by Kugel and 
Khomskii many years ago \cite{kugel} to understand the magnetic properties 
of transition metal oxides which have orbital degeneracies. Recently physical 
realizations of the spin-1/2 Kitaev model have been proposed in optical 
lattice systems \cite{demler} and in quantum circuits \cite{nori}. Variants 
of the model have also been studied in two dimensions 
\cite{wen,yao,yang,dusuel,kells,yao2,wu,kamfor,shi}, three dimensions 
\cite{si,mandal} and
also on quasi-one-dimensional lattices \cite{karimipour,sen,saket}. Finally, 
the spin-$S$ Kitaev model has been studied in the large $S$ limit using spin 
wave theory \cite{baskaran2}, and the classical version of the Kitaev model 
has been studied at finite temperatures using analytical and Monte Carlo 
techniques \cite{chandra}. Their results indicate that while the phenomenon of
order-by-disorder \cite{villain,moessner,henley,gvozdikova} may occur in the 
quantum mechanical Kitaev model, it does not in the corresponding classical 
model. 

For the Kitaev model with spin $S > 1/2$, there is a $\mathbb{Z}_2$ 
invariant associated with each plaquette for arbitrary spin-$S$, which 
reduces to the conserved $\mathbb{Z}_2$ gauge flux for the spin-1/2 case 
\cite{baskaran2}. However, the model does not seem to be fully integrable. 
While some differences in the structure of the invariants between the models 
with half-odd-integer and integer spins have been pointed out 
\cite{baskaran2}, the issue of whether
there are systematic differences in the nature of the low-energy spectrum is 
also of interest. In the present paper, we approach this problem by examining 
the spin-1 Kitaev model. The two-dimensional model appears difficult to 
analyze, but even the one-dimensional version of it has a lot of interesting 
structure, as we proceed to show.

The plan of this paper is as follows. In Sec. \ref{1dkm}, we consider the 
spin-$S$ Kitaev chain. In Sec. \ref{inv} we show that this model has local, 
mutually commuting conserved quantities $W_n$, for integer $S$. The eigenvalues
of $W_n$ are $\pm 1$. For open boundary conditions, there are some additional 
conserved quantities at the ends of the system. The existence of these 
conserved quantities implies that the Hilbert space of a $N$-site system can 
be decomposed into a sum of $2^N$ disjoint subspaces. The dimensions of these 
subspaces are not equal. In Sec. \ref{count} we develop a formalism to 
compute the dimension of these sectors. For large $N$, the dimension varies 
as $d^N$ in most sectors, with the constant $d$ depending on the sector. The 
sectors show complicated spatial structures, arising from the spatial 
structure of $\{W_n\}$. We show this in Sec. \ref{spinev}, by computing the 
non-trivial spatial dependence of expectation values of spin operators in 
some sectors, averaged over all states in the sector. 
We then consider the spin-1 model in Sec. \ref{sp1}. In Sec.
\ref{num}, we consider the ground state and lowest excited state of the system.
Exact diagonalizations of small systems show that the ground state lies in a
sector in which $W_n=+1$ for all $n$. In this sector, there is a gap
between the ground state and the first excited state. The lowest excited state
of the system is the ground state of a different sector; and the energy-gap 
seems to approach a non-zero value in the limit of the system size going to 
infinity. In Sec. \ref{shalf}, we consider the sector containing the ground 
state, and show that the Hamiltonian is equivalent to the Hamiltonian of a 
deposition-evaporation process of a nearest-neighbor-exclusion lattice gas 
model, which can be written as of a spin-$1/2$ system with local interactions 
with a range extending to at most next-nearest neighbors. The Hamiltonian 
seems to be difficult to diagonalize exactly, we present a variational study 
of the ground state in Sec. \ref{var}. The variational estimate
of the ground state energy is found to agree well with the results obtained
numerically for small systems. We also analyze the first excited state of the
Hamiltonian. In Sec. \ref{dilute}, we consider a more general Hamiltonian, 
obtained by adding a term $\lambda \sum_n W_n$, and 
discuss its ground states as a function of $\lambda$. We show that the ground 
state of this new Hamiltonian is gapless for a range of couplings 
$\lambda^c_1 \leq \lambda \leq \lambda^c_2$, and gapped otherwise. We argue 
that for $\lambda$ just above $\lambda^c_1$, in the sector containing the 
ground state, the density of negative $W$'s is of order $|\frac{1}{\log 
(\lambda - \lambda^c_1)}|$. For $\lambda$ just below $\lambda^c_2$, 
the density of positive $W$'s goes to zero as $(\lambda^c_2 - 
\lambda)^{1/2}$. In the final section, we summarize our conclusions, and 
discuss the relationship of this model with the Fibonacci chain.

\section{One-dimensional Kitaev model} 
\label{1dkm}

In this section, we will discuss a one-dimensional spin-$S$ model which is 
obtained by considering a single row of the Kitaev model in two dimensions.

Let us begin with the Kitaev model on the honeycomb lattice. This is governed 
by the Hamiltonian
\beq H_{Kit}^{hex}=J_x\sum_{\langle ij\rangle_x}S^x_iS^x_j +J_y\sum_{\langle ij
\rangle_y} S^y_iS^y_j +J_z\sum_{\langle ij\rangle_z}S^z_iS^z_j, \eeq
where $\langle ij\rangle_a$ denote the nearest-neighbor bonds in the 
$a^{\rm th}$ direction. If we set $J_z=0$, we get a set of decoupled chains. 
We call this the Kitaev chain, and this is the topic of this paper. The 
Hamiltonian is by
\beq H=\sum_n\left( J_{2n-1} S^x_{2n-1}S^x_{2n}+ J_{2n} S^y_{2n}S^y_{2n+1}
\right). \label{ham1} \eeq
In general, the couplings $J_m$ could be all different from each other.
If some of the couplings are negative, we can change the signs of those
couplings by performing the unitary transformation
\beq S^x_m \to - S^x_m, ~~S^y_m \to - S^y_m, ~~{\rm and} ~~S^z_m \to S^z_m \eeq
on appropriate sites. We consider the simpler case, where all couplings have 
the same value,
$J_m = J$. Without any loss of generality, we set $J= 1$. Finally, the 
Hamiltonian can be unitarily transformed to a more convenient form by the 
following transformation on the even sites, 
\beq S^x_{2n} \to S^y_{2n}, ~~S^y_{2n} \to S^x_{2n}, ~~{\rm and} ~~S^z_{2n} 
\to - S^z_{2n}. \eeq
The Hamiltonian in Eq. (\ref{ham1}) then takes the translation invariant form
\beq H_{Kit}=\sum_n~S^x_nS^y_{n+1}. \label{ham2} \eeq

\begin{figure}[t] {\includegraphics*[width=\linewidth]{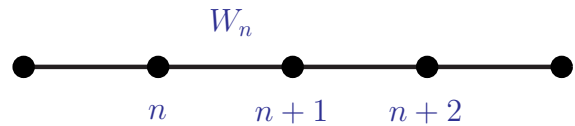}}
\caption{Picture of the Kitaev chain showing one of the conserved quantities
$W_n$.} \label{fig1} \end{figure}

\subsection{Invariants}
\label{inv}

The Hamiltonian in Eq. (\ref{ham2}) has the following local symmetries for all
$S$. Let us introduce the operators on sites
\beq \Si^a_n=e^{i\pi S_n^a}, \eeq 
and operators on bonds
\beq W_n ~=~ \Si^y_n \Si^x_{n+1}. \label{cons} \eeq
as shown in Fig. 1. We then find that
\beq [W_n, H]=0. \eeq
The eigenvalues of $\Si^a_n$ are $\pm 1$ for integer $S$ and
$\pm i$ for half-odd-integer $S$. Thus for any value of the spin $S$, the
eigenvalues of $W_n$ are $\pm 1$. 

However, there is a qualitative difference between integer and 
half-odd-integer values of $S$. For integer values of $S$, all the matrices 
$\Si^a_n$ matrices commute with each other, whereas for half-odd-integer 
values, $\Si^a_n$ commutes with $\Si^b_m$ for $n\ne m$ but anticommutes with 
$\Si^b_n$ for $a\ne b$. Consequently, for integer $S$, all the invariants 
$W_n$ commute, but for half-odd-integer $S$, $W_n$ anti-commutes with its 
neighboring invariants, $W_{n\pm 1}$, and commutes with $W_m$, $m\ne n,n\pm 
1$. We will now show that this implies that all the eigenstates of the chain 
with half-odd-integer $S$ are $2^{N/2}$ fold degenerate. 

The invariants for half-odd-integer $S$ can be combined in the following way 
to form a set of mutually commuting angular momentum operators, one per 
every two bonds,
\beq \mu^z_n=W_{2n},~~ \mu^x_n=W_{2n-1}\prod_{m<n}W_{2m-1},~~
\mu^y_n=i\mu^x_n\mu^z_n. \eeq
It can be verified that
\bea \label{mualg1} \left[\mu_n^a,\mu^b_m\right]&=& 2i\delta_{nm}\ep^{abc}
\mu^c, \\
\label{mualg2} \left\{\mu_n^a,\mu^b_n\right\}&=&2\delta^{ab}. \eea
The $\mu_n^a$ commute with the Hamiltonian as they are made by multiplying 
conserved operators. Hence Eq. (\ref{mualg1}) shows that the Hamiltonian
has a $\left(SU(2)\right)^{N/2}$ symmetry, where $N$ is the number of sites. 
Eq. (\ref{mualg2}) shows that each of these $SU(2)$ factors are realized 
in the spin-$1/2$ representation. Thus each eigenstate is $2^{N/2}$-fold 
degenerate. There is no reason for such a degeneracy for integer $S$ and 
indeed, as we will see later, the ground state for $S=1$ is non-degenerate.

We note that the spin-$S$ Kitaev model in two dimensions also has a 
$\mathbb{Z}_2$ valued invariant associated with every hexagon of the 
honeycomb lattice \cite{baskaran2}. When they are restricted to a single 
chain, the invariants take the form 
\beq V_n ~=~ \Si^y_n \Si^z_{n+1} \Si^x_{n+2} \label{cons2} \eeq
which involves three neighboring sites. The invariants given in Eq. 
(\ref{cons}) are simpler because they only involve two sites. For any spin 
$S$, we find that $\Si^x_n \Si^y_n \Si^z_n = I$ and $(\Si^a_n)^2 = (-1)^{2S}$;
hence the invariants in Eqs. (\ref{cons}) and (\ref{cons2}) are related to 
each other as
\beq V_n ~=~ (-1)^{2S} ~W_n W_{n+1}. \eeq

Open chains have some extra symmetries at the edges. If the site labels of the 
open chain are $1,\cdots,N$, then $S^x_1$ and $S^y_N$ also commute with the
Hamiltonian. Thus at the first and last sites, we have a $U(1)$ symmetry group
generated by these operators. Note that a $\mathbb{Z}_2$ subgroup of this 
group, consisting of the operators $\Si^x_1$ and $\Si^y_N$, also commutes with
all the invariants. If we combine the operators $S^x_1$ and $S^y_N$ with the 
$W_n$ invariants on the first and last bonds, we have a larger symmetry group 
made of $W_1,~S^x_1$ and their products at the first bond, and the group made
of $W_N,~S^y_N$ at the last bond. As we will show in Sec. \ref{sp1}, for 
the $S=1$ case the group formed is $SU(2) \times U(1)$ at each end. 

\subsection{Counting of states for integer $S$}
\label{count}

We will now develop a formalism to count the number of states in a given
sector for integer $S$. In this case, the $\Sigma$ matrices commute and 
hence can be simultaneously diagonalized. If $\vert S,m\rangle$ denote the 
eigenstates of $S^z$, then 
\bea \label{sigxac} \Sigma^x\vert S,m\rangle&=&(-1)^S\vert S,-m\rangle, \\
\label{sigyac} \Sigma^y\vert S,m\rangle&=&(-1)^{S+m}\vert S,-m\rangle, \\
\label{sigzac} \Sigma^z\vert S,m\rangle&=&(-1)^m \vert S,m\rangle. \eea
We can construct the eigenstates of the $\Sigma$ matrices in the $m\ne 0$
subspace,
\beq \label{mpmdef} \vert S,m\pm\rangle\equiv \frac{1}{\sqrt{2}}\left(\vert 
S,m\rangle\pm\vert S,-m\rangle\right), \eeq
where $m=1,\cdots,S$. The eigenvalues of the matrices are,
\bea \label{sigxev} \Sigma^x\vert S,m\pm\rangle&=&\pm(-1)^S\vert S,m\pm
\rangle, \\
\label{sigyev} \Sigma^y\vert S,m\pm\rangle&=&\pm(-1)^{S+m}\vert S,m\pm
\rangle, \\
\label{sigzev} \Sigma^z\vert S,m\pm\rangle&=&(-1)^m \vert S,m\pm\rangle. \eea

The states of a chain can be classified by the eigenvalues of $\Sigma^x_n$ 
and $\Sigma^y_n$ as 
\beq (y_Nx_N)\cdots(y_2x_2)(y_1x_1), \eeq
where $x_n,y_n=\pm 1$ are the eigenvalues of $\Sigma^x_n$ and $\Sigma^y_n$ 
respectively. The invariants are then $W_n=x_{n+1}y_n$.

We now calculate the number of states in a given sector ${\cal W}$ using a 
standard transfer matrix technique. Consider the allowed states of $r$ sites, 
when the values of $r-1$ constants $W_j$, with $j = 1$ to $r-1$ have been 
specified. We denote this set of values by ${\cal W}$. Let $Z_r(y |{\cal W})$ 
denote the number of allowed states of this set of sites with $\Sigma^y_r = y$
where $y$ takes values $\pm 1$. We now add a site $r+1$ to the chain, and 
also specify $W_r$. Let the new set of $\{W\}$ be denoted by ${\cal W}'$.

Consider first the case $W_r = +1$. Clearly, we can have two possibilities: 
$\Sigma^x_{r+1} = \Sigma^y_r = +1$, or $\Sigma^x_{r+1} = \Sigma^y_r = -1$. 
Let $\nu(p,p')$ denote the number of states of a single site with $\Sigma^y 
= p$, and $\Sigma^x = p'$. Then, we clearly have the recursion equation
\beq Z_{r+1}(y|{\cal W}') = \nu(y,+1) Z_r(+1,{\cal W}) + 
\nu(y, -1) Z_r(-1,{\cal W}). \eeq
This equation can be written as a matrix equation
\bea \left[\begin{array}{l}Z_{r+1}(+1|{\cal W}')\\Z_{r+1}(-1|{\cal W}')
\end{array}\right] = {\mathbb T}_+ {\left[\begin{array}{l}Z_r(+1|{\cal W})\\
Z_r(-1|{\cal W})\end{array}\right]}, \eea
where ${\mathbb T}_+$ is a $2 \times 2$ matrix given by
\bea {\mathbb T}_+&=& \left[\begin{array}{ll}\nu(+1,+1)&\nu(+1,-1)\\
\nu(-1,+1)&\nu(-1,-1) \end{array}\right], \eea
It then follows from Eqs. (\ref{sigxev}-\ref{sigzev}) that
\bea \label{tplus} {\mathbb T}_+&=& \frac{1}{2}\left[\begin{array}{ll}S-1&S+1\\
S+1&S+1 \end{array}\right] ~~{\rm for}~S~{\rm odd}, \\
\label{nueven} &=& \frac{1}{2}\left[\begin{array}{ll}S+2&~~~S~~\\ S&~~~S~~
\end{array}\right] ~~{\rm for}~S~{\rm even}. \eea

Similarly, when $W_r = -1$, the corresponding recursion equation is
\bea \left[\begin{array}{l}Z_{r+1}(+1|{\cal W}') \\ Z_{r+1}(-1|{\cal W}')
\end{array}\right] = {\mathbb T}_- {\left[\begin{array}{l}Z_r(+1|{\cal W}) \\
Z_r(-1|{\cal W})\end{array}\right]}, \eea
where the matrix ${\mathbb T}_-$ is given by
\beq {\mathbb T}_- = {\mathbb T}_+ \tau^x, {\rm ~~~~with} ~~~~~\tau^x = 
\left[\begin{array}{ll}0 ~~1\\1 ~~0 \end{array}\right]. \eeq

It is then clear that for a given set of invariants ${\cal W}$, the number of 
states can be written in terms of a product of the matrices ${\mathbb T}_+$ 
and ${\mathbb T}_-$. 

For example, for an open chain of $N$ sites, and
${\cal W} = \{W_{N-1},...W_3, W_2, W_1\}=\{+1,...+1,-1,-1\}$, we have
\bea \left[\begin{array}{l}Z_N(+1|...+--)\\Z_N(-1|...+--)\end{array}\right] 
= {\mathbb T}_+...{\mathbb T}_+ {\mathbb T}_- {\mathbb T}_- {\left[
\begin{array}{l}Z_1(+1|\phi)\\Z_1(-1|\phi)\end{array}\right]}, \eea 
where $\phi$ denotes the null string, and $Z_1(y|\phi)$ denotes the number of 
states of the spin at site $1$ with $\Sigma_1^y = y$. Thus $Z_1(+1|\phi) 
= S+1$, $Z_1(-1|\phi) = S$, when $S$ is an even integer, and $Z_1(+1|\phi) = 
S$, $Z_1(-1|\phi) = S + 1$ when $S$ is an odd integer. The total number of 
states in this sector is then given by
\beq \Gamma({\cal W}) = Z_N(+1|{\cal W}) + Z_N(-1|{\cal W}). \eeq

For a closed chain, there is an additional invariant $W_N = y_N x_1$ 
and the number of states in the sector becomes
\beq \Gamma({\cal W})={\rm Tr}\left(\prod_{n=1}^N {\mathbb T}_{W_n}\right), 
\label{closedstates} \eeq
where ${\mathbb T}_{W_n} \equiv {\mathbb T}_{\pm}$ for $W_n = \pm 1$ and 
$\prod_{n=1}^N$ is an ordered product of ${\mathbb T}_{\pm}$ matrices, from 
site $1$ to $N$ with the index increasing from right to left.

We now calculate the dimensions of some sectors for a closed chain of length 
$N$. It is easy to get an explicit answer for the two extreme limits when 
$W_n=\pm 1$ for all $n$. In these cases, the number of states, $\Gamma^\pm$, is
\beq \label{gammapmsol} \Gamma^\pm=\left(d_1^\pm\right)^N +\left(
d_2^\pm\right)^N , \eeq
where $d_1^\pm(S)$ and $d_2^\pm(S)$ are the larger and smaller eigenvalues of 
${\mathbb T}_{\pm}$ respectively. The eigenvalues can be computed to give,
\bea \label{dpodd}
d^+_{1(2)}&=&\frac{1}{2}\left(S\pm\sqrt{S^2+2S+2}\right)~~{\rm for}~S~
{\rm odd}, \\ \label{dpeven} 
&=&\frac{1}{2}\left(S+1\pm\sqrt{S^2+1}\right)~~{\rm for}~S~{\rm even}, \\ 
\label{dmodd}
d^-_{1(2)}&=&\frac{1}{2}\left(S+1\pm\sqrt{S^2-1}\right)~~{\rm for}~S~
{\rm odd}, \\ \label{dmeven}
&=&\frac{1}{2}\left(S\pm\sqrt{S^2+2S}\right)~~{\rm for}~S~{\rm even}. \eea

For $S=1$, $d^+_1$ is equal to the golden ratio, $\gamma=(1+\sqrt{5})/2$, and
$d^+_2=-1/\gamma$. As $N \to \infty$, the dimension of the Hilbert space in 
the sector with all $W_n=1$ grows as $\gamma^N$. On the other hand, $d^-_1 =
d^-_2=1$. The dimension of the sector with all $W_n=-1$ is therefore equal to 
2.

With the exception of $S=1$, the larger of the two eigenvalues $d^\pm_1$ is
always greater than 1, and in the $N\to\infty$ limit, we have
\beq \label{qdans} \Gamma^\pm(S)=\left(d_1^\pm(S)\right)^N. \eeq
$d_1^\pm(S)$ is referred to as the quantum dimension of the sector. As can 
be seen it is, in general, fractional for any $S$. In the limit $S \to \infty$,
the quantum dimension tends to $S+1/2$ for both the sectors. It is interesting
to note that it is a half-odd-integer in this limit.

\subsection{Expectation values of the $\Sigma$ operators in different sectors}
\label{spinev}

In this section we find the expectation values of the $\Sigma_n^a$ operators in
various sectors. We will assume periodic boundary conditions. Our calculation 
will average over all the states of a given sector considered with equal 
weight; this can be considered as a calculation in the limit that the 
temperature $T \to \infty$, so that it does not depend on the Hamiltonian.

We evaluate the expectation values of $\Sigma_n^a$ by inserting projection 
operators at site $n$ in the product of transfer matrices in Eq. 
(\ref{closedstates}). This yields the following expression for the expectation 
value of the $\Sigma_n^a$ operator in a general sector with a 
$W$-configuration ${\cal W}$

\begin{align}
\non &\langle \Sigma_{n+1}^a \rangle_{{\cal W}} = {\rm Tr}\left[\left(
\prod_{j=n+1}^N {\mathbb T}_{W_j}\right){\mathbb T}_{W_n}^a\left(
\prod_{i=1}^{n-1} {\mathbb T}_{W_i}\right)\right]/\Gamma(\{{\cal W}\})\\
\non & \textmd{where}\\
\non &{\mathbb T}_{W_n}^x = W_n {\mathbb T}_{W_n} \tau^z\\
\non &{\mathbb T}_{W_n}^y = \tau^z {\mathbb T}_{W_n}\\
&{\mathbb T}_{W_n}^z = W_n \tau^z {\mathbb T}_{W_n} \tau^z \end{align}
and $\tau^z$ and $\tau^x$ are the well-known Pauli matrices. 

We now compute the expectation values of $\Sigma_n^a$ in two sectors: 
the sector ${\cal W}_0$ with all $W_n =+1$, and the sector ${\cal W}_1$ 
in which one of the $W_n =-1$ and all the other $W_n =+1$ (without loss of 
generality we pick $W_N = -1$). The expressions for $\langle \Sigma_n^a
\rangle_{{\cal W}_{0,1}} \equiv \langle \Sigma_n^a\rangle_{0,1}$ can be 
evaluated in terms of the eigenvectors and eigenvalues of ${\mathbb T}_+$. 
The ${\mathbb T}_+$ matrix is a linear combination of the Pauli matrices, 
$\tau^z$ and $\tau^x$. Its eigenvectors are spinors polarized parallel and 
anti-parallel to a direction in the $z_x$ plane, forming an angle $\theta_S$ 
with the $z-$axis, where $\theta_S$ is defined by,
\bea \non
\cos\theta_S&\equiv&-\frac{1}{\sqrt{1+(S+1)^2}}~~{\rm for}~S~{\rm odd}, \\
\label{cosdef} &\equiv&\frac{1}{\sqrt{1+S^2}}~~{\rm for}~S~{\rm even}, \\
\non \sin\theta_S&\equiv&\frac{S+1}{\sqrt{1+(S+1)^2}}~~{\rm for}~S~
{\rm odd}, \\ \label{sindef}
&\equiv&\frac{S}{\sqrt{1+S^2}}~~{\rm for}~S~{\rm even}. \eea

For the sector with all $W_n=+1$ it is easy to see that $\langle\Sigma_n^x
\rangle_0=\langle\Sigma_n^y\rangle_0$. For large $N$, we obtain
\bea \label{sigxyans}
&\langle\Sigma^{x(y)}_n\rangle_0=\cos\theta_S, \\ \label{sigzans}
&\langle\Sigma^z_n\rangle_0=\cos^2\theta_S+\frac{d^+_2}{d^+_1}\sin^2\theta_S.
\eea

In the sector where $W_N=-1$ and the rest are equal to $+1$, we get, for 
large $N$,
\bea \label{sigxexpminus2}
\langle\Sigma^x_n\rangle_1&=& \langle\Sigma^x_n\rangle_0
\left(1-\left(\frac{d^+_2}{d^+_1}\right)^{n-1}\right), \\
\label{sigyexpminus2}
\langle\Sigma^y_n\rangle_1&=& \langle\Sigma^y_n\rangle_0
\left(1-\left(\frac{d^+_2}{d^+_1}\right)^n\right), \\ \non
\langle\Sigma^z_n\rangle_1&=& \langle\Sigma^z_n\rangle_0
\left(1+\frac{2\cos\theta_S}{d^+_1\cos^\theta_S+d^+_2\sin^2\theta_S}
\left(\frac{d^+_2}{d^+_1}\right)^{n-1}\right). \\ \label{sigzexpminus2}
&& \eea

Note that in the limit $n \to \pm \infty$, $\langle \Sigma_n^a \rangle$ in 
Eqs. (\ref{sigxexpminus2}-\ref{sigzexpminus2}) approach the values given 
in Eqs. (\ref{sigxyans}-\ref{sigzans}) exponentially quickly.

\begin{figure} \includegraphics[scale=0.58,angle=0]{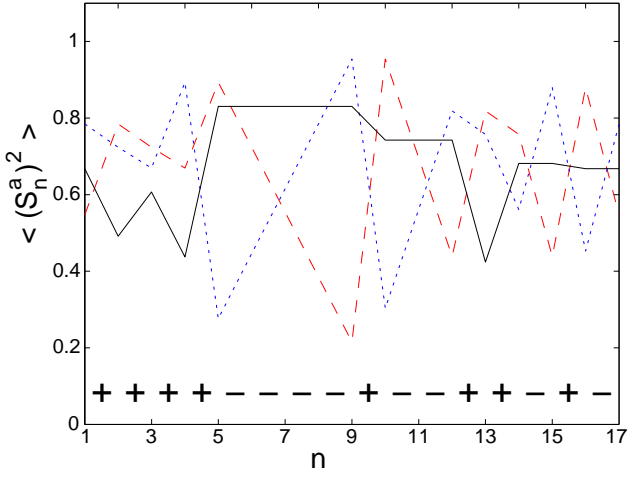} 
\caption{Plot of $\langle{S^x_n}^2\rangle$ (dotted line), $\langle{S^y_n}^2
\rangle$ (dashed line) and $\langle{S^z_n}^2\rangle$ (full line) as a function
of $n$ for $S=1$, on a ring with $16$ sites in the sector $++++----+--++-+-$, 
with periodic boundary conditions (site 17 = site 1).} \label{fig:expect} 
\end{figure}

While in general $\Sigma_n^a$ are complicated multi-spin operators, for 
$S=1$ we have $\Sigma_n^a=1-2(S_n^a)^2$. Thus, for $S=1$ we are essentially 
computing the expectation values of $(S_n^a)^2$. To see what the spin textures 
are like in a typical sector, we have plotted in
Fig. \ref{fig:expect} the expectation values of ${S_n^x}^2, {S_n^y}^2$ and
${S_n^z}^2$ for $S=1$, as a function of the spatial coordinate $n$ for a ring 
of size $16$, in the sector where the sequence of $W$'s is $++++----+--++-+-$.
This sequence was chosen as it is a de Bruijn sequence \cite{bruijn} of length
16, in which each of the 16 possible binary sequences of length 4 occur exactly
once, taking the periodic boundary conditions into account.

\section{$S=1$ model}
\label{sp1}

We will now focus on the Kitaev chain with spin-1's at each site.
We will work with the natural spin-1 representation in which
\beq \left(S^a\right)_{bc}=i\ep_{abc}. \label{s1adef} \eeq
In this representation, the matrices $\Si^a$ are diagonal and are given by
\bea\label{sigx1} \Si^x &=& \left( \begin{array}{lll}
1&0&0 \\
0&-1&0 \\
0&0&-1 \end{array} \right), \non \\
\Si^y &=& \left( \begin{array}{lll}
-1&0&0 \\
0&1&0 \\
0&0&-1 \end{array} \right), \non \\
\Si^z &=& \left( \begin{array}{lll}
-1&0&0 \\
0&-1&0 \\
0&0&1 \end{array} \right). \label{Sixyz} \eea
We note that these matrices satisfy $\Si^x \Si^y \Si^z = I$. We denote the 
basis vectors by $\vert x\rangle,~\vert y\rangle$ and $\vert z\rangle$ 
defined as
\beq \label{xyzketdef} 
\vert x \rangle = \left( \begin{array}{c}
1 \\ 0 \\ 0 \end{array} \right),~~~
\vert y \rangle=\left(\begin{array}{c}
0 \\ 1 \\ 0 \end{array} \right),~~~
\vert z \rangle=\left(\begin{array}{c}
0 \\ 0 \\ 1 \end{array} \right). \eeq

We then see that the 9 possible states at sites $(n,n+1)$ are given by:
\beq |xy\rangle, ~|xz\rangle, ~|yx\rangle, ~|zy\rangle~~ {\rm and} ~~|zz
\rangle ~~{\rm with}~~ W_n = 1, \label{wplus} \eeq
and 
\beq |xx\rangle, ~|yy\rangle, ~|yz\rangle~~ {\rm and} ~~|zx\rangle ~~
{\rm with} ~~W_n = -1. \label{wminus} \eeq 
{}From Eq. (\ref{s1adef}) we have,
\bea S^x\vert x\rangle &=& 0,~~~~~~~~~~S^y\vert x\rangle=i\vert z\rangle,~~~~~~
S^z\vert x \rangle=-i\vert y\rangle, \non \\
S^x\vert y\rangle &=& -i\vert z\rangle,~~~~S^y\vert y\rangle=0,~~~~~~~~~ S^z
\vert y \rangle=i\vert x\rangle, \non \\
S^x\vert z\rangle &=& i\vert y\rangle,~~~~~~S^y\vert z\rangle=-i\vert x
\rangle,~~~~S^z\vert z\rangle=0. \label{sxyz} \eea
Eqs. (\ref{Sixyz}) and (\ref{sxyz}) imply that $(S^a)^2 = (1 - \Si^a)/2$.

For the 5 states in Eq. (\ref{wplus}) satisfying $W_n = 1$, we have the 
following actions of the relevant term in the Hamiltonian,
\bea S^x_1S^y_2\vert xy\rangle &=& 0, \non \\
S^x_1S^y_2\vert xz\rangle &=& 0, \non \\
S^x_1S^y_2\vert zy\rangle &=& 0, \non \\
S^x_1S^y_2\vert zz\rangle &=& \vert yx\rangle, \non \\
S^x_1S^y_2\vert yx\rangle &=& \vert zz\rangle. \label{sxsy1} \eea
For the 4 states in Eq. (\ref{wminus}) satisfying $W_n = -1$, the actions of 
the relevant term in the Hamiltonian are given by
\bea S^x_1S^y_2\vert xx\rangle &=& 0, \non \\
S^x_1S^y_2\vert yy\rangle &=& 0, \non \\
S^x_1S^y_2\vert yz\rangle &=& -\vert zx\rangle, \non \\
S^x_1S^y_2\vert zx\rangle &=& -\vert yz\rangle. \label{sxsy2} \eea

As mentioned earlier, for an open chain with site numbers going from 1 to $N$,
we find that $S_1^x$ and $S_N^y$ commute with $H$. We define the operators,
\bea \label{tau12def} \tau^1\equiv iW_1S^x_1, &~~~&\tau^2\equiv S^x_1, \\
\label{tau30def} \tau^3 \equiv -S^x_1W_1S^x_1, &~~~& \tau^0\equiv 
\frac{1}{2}\left(1-(S^x_1)^2\right). \eea
It can be verified that these operators obey a $SU(2)\times U(1)$
algebra. Exactly the same construction on the last bond, with $S^x_1
\to S^y_N$ and $W_1\to W_N$, yields the same algebra on that bond.

\subsection{Numerical studies}
\label{num}

\begin{figure} \includegraphics[scale=0.68]{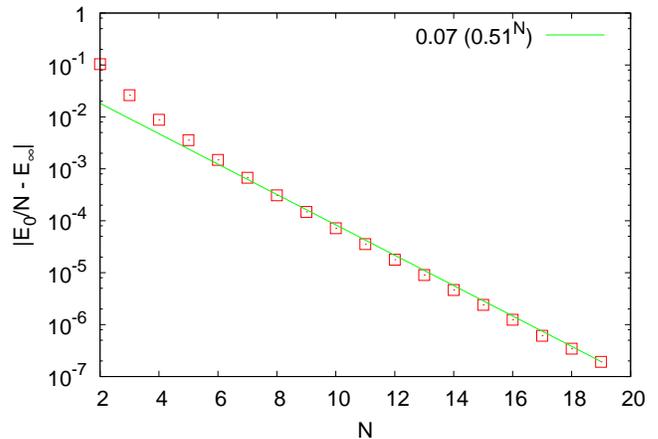}
\caption{Graph of $|E_0 /N - E_{\infty}|$ with $N$, where $E_{\infty} = 
-0.60356058$, showing an exponential convergence to the value of the ground 
state energy with $N$.} \label{eigenvaluesfit} \end{figure}

We have carried out exact diagonalization studies of small systems with
periodic boundary conditions in order to find the energies of the ground state 
and the lowest excited state of the spin-1 Kitaev chain. We find that the 
ground state lies in the sector with all $W_n = 1$ and has zero momentum 
(momentum is a good quantum number 
in this sector since the values of the $W_n$'s are translation invariant). 
The ground state energy per site as a function of the system size 
$N$ is presented in Table I. We see that $E_0/N$ shows odd-even oscillations
as a function of $N$ but seems to converge quite fast. The fast convergence 
indicates that the ground state must have a fairly short correlation length.
The $N$-dependence of $\bar{E}_N = E_0/N$ can be fitted to the form 
\beq \bar{E}_N = E_{\infty} + B (-\alpha)^N. \eeq
A simple plot of $\log| \bar{E}_N - E_{\infty}|$ versus $N$ (Fig. 
\ref{eigenvaluesfit}), gives a good straight line for $E_{\infty} =
-0.60356058$, which we take to be the best estimate of $E_{\infty}$. The 
corresponding values of $B$ and $\alpha$ are $0.07$ and $0.51$. The 
estimated errors of extrapolation are about $1$ in the last significant digit. 

\begin{table}[h] 
\begin{center} \begin{tabular}{|c|c|c|c|} \hline
$N$ & $E_0/N$ & $N$ & $E_0/N$\\ \hline
2 & -0.707106770 & 11 & -0.603525102 \\ \hline
3 & -0.577350259 & 12 & -0.603578389\\ \hline
4 & -0.612372458 & 13 & -0.603551567\\ \hline
5 & -0.600000024 & 14 & -0.603565216 \\ \hline
6 & -0.605046094 & 15 & -0.603558183 \\ \hline
7 & -0.602888465 & 16 & -0.603561819\\ \hline
8 & -0.603869855 & 17 & -0.603559971 \\ \hline
9 & -0.603412688 & 18 & -0.603560924 \\ \hline
10 & -0.603632331& 19 & -0.603560388 \\ \hline
\end{tabular} \end{center}
\caption{Ground state energy per site versus $N$.} 
\end{table}

In the sector with all $W_n = 1$,
the first excited state has momentum equal to $\pi$ if $N$ is even. We find 
that the gap separating it from the ground state is given by $1.0353$ for 
$N=4$ and $0.9845$ for $N=6$. These values also seem to be converging rapidly,
and the large value is consistent with a short correlation length. 
However, this is {\it not} the lowest excited state of the system. Rather, we 
find that the state nearest in energy to the ground state is the ground state 
of the sector with exactly one $W_n = -1$ and all the other $W_n =1$.
(We cannot use momentum to classify the states in this sector since it is not 
translation invariant). The energy gap $\Delta E$ between the lowest energy 
state in this sector and the ground state of the sector with all $W_n = 1$ is 
shown in Table II. We see that these also oscillate between even and 
odd values of $N$ but seem to converge quite fast to a small but non-zero 
value. This is evidence that the spin-1 Kitaev chain has a finite gap 
in the thermodynamic limit $N \to \infty$.

\begin{table}[h] 
\begin{center} \begin{tabular}{|c|c|} \hline
$N$ & $\Delta E$ \\ \hline
3 & 0.1141 \\ \hline
4 & 0.2025 \\ \hline
5 & 0.1671 \\ \hline
6 & 0.1802 \\ \hline
\end{tabular} \end{center}
\caption{Energy gap between ground state and first excited state versus $N$.}
\end{table}

\section{Mapping the spin-1 chain to a spin-1/2 chain}
\label{shalf}

For a given value of the state of
the spin at site $n$, and a given value of $W_n$, there are at most two
choices for the spin state at site $n+1$. Hence it is clear that the 
Hilbert space of a given sector can be mapped into the Hilbert space
of a spin-1/2 chain, with some states excluded which correspond to
infinite energy. However, in general, the corresponding Hamiltonian would 
have a rather complicated form, with long-ranged interactions. The mapping 
is easy to construct explicitly in the sector with all $W_n = +1$, and 
the corresponding Hamiltonian has only local interactions. This is what we 
now proceed to show.

Consider the state $zzzz\cdots$ that belongs to the sector with all $W_n = 
+1$. The only allowed process in this sector is $zz \rightleftharpoons
yx$ [Eqs. (\ref{sxsy1})]. We may think of this process as a quantum dimer
deposition-evaporation model. The $z$-spins are treated as empty sites;
two empty sites can be changed to being occupied by a dimer $yx$ by a
`deposition' process, and conversely, $yx$ can `evaporate' and become
$zz$ again. The dimers have a hard-core constraint, and a site cannot
be shared by two dimers. The dimers are oriented: the `head' $x$ being 
to the right of the `tail' $y$.

This dimer deposition-evaporation model can also be described as a
deposition-evaporation of a nearest-neighbor exclusion lattice gas. We
just think of the heads as particles, and do not distinguish between the
tails and empty sites, except for ensuring that we deposit a particle at a
site only if it is empty and both its nearest neighbors are also empty.
Then this model is described by the Hamiltonian
\beq H_d ~=~ -~ \frac{1}{4} ~\sum_n~ (1 - \si^z_{n-1}) ~\si^x_n (1 - 
\si^z_{n+1}). \label{ham3} \eeq

We note that this model is different from the dimer
deposition-evaporation models studied earlier \cite{dde}, in that the
two ends of the dimer are distinct, and there is no reconstitution.
Also, this Hamiltonian does not have an interpretation as the evolution
operator of a classical Markov process, as there are no diagonal terms
corresponding to probability conservation.

We have introduced a minus sign in the Hamiltonian for later convenience. 
This does not change the eigenvalue spectrum as the eigenvalues of $H_d$ 
occur in pairs $\pm e_i$.

\section{Variational study of sector with all $W_n = 1$}
\label{var}

We will now use a variational approach to study the ground state of the 
Hamiltonian $H_d$ with periodic boundary conditions.
We use the $z$-basis, and denote the $\up$ state at the site $i$ by an 
occupied site ($n_i =1$), and the $\dn$ state by an empty state ($n_i =0$).
Since two adjacent sites cannot be simultaneously occupied, the state space 
is that of hard-core particles with nearest-neighbor exclusion on a line. A 
configuration ${\bf C}$ is specified by an $N$-bit binary string $0010010101
\cdots$, which gives the values of all the $N$ occupation numbers $n_i$. We 
note that in the basis where all the $n_i$ are diagonal, the Hamiltonian $H_d$
has all matrix elements non-positive. This implies that the (real) eigenvector
corresponding to the lowest energy will have all components of the same sign 
in this basis. 

For the ground state of $H_d$, we consider a variational wave function of the 
form 
\beq |\psi\rangle = \sum_{\bf C} \sqrt{{\rm Prob}({\bf C})}~ |{\bf C}\rangle,
\eeq
where ${\rm Prob}({\bf C})$ is chosen as the probability of the lattice gas 
configuration ${\bf C}$ in some classical equilibrium ensemble corresponding 
to a suitably chosen lattice gas Hamiltonian. Clearly, this trial vector is 
normalized, with 
\beq \langle \psi | \psi \rangle = 1. \eeq
With this choice, ${\rm Prob}({\bf C})$ is also the probability of the 
configuration ${\bf C}$ in the quantum mechanical variational state 
$|\psi\rangle$.

The simplest choice of the lattice-gas Hamiltonian is that of a classical 
lattice gas with nearest-neighbor exclusion, and a chemical potential $\mu$, 
with a Hamiltonian given by
\beq H_{cl} = + \infty \sum_i n_i n_{i+1} - \mu \sum_i n_i, \label{ham4} \eeq
where we use the convention that $0\cdot\infty=0$; hence the first term in Eq.
(\ref{ham4}) allows states with $n_i n_{i+1} = 0$ but disallows states with 
$n_i n_{i+1} = 1$. Let us denote $z = \exp(\beta \mu)$. It is straightforward 
to determine various correlation functions in the thermal equilibrium state 
corresponding to $H_{cl}$. The probability of a configuration ${\bf C}$ 
is given by 
\beq {\rm Prob}({\bf C}) = \exp[ -\beta H_{cl}({\bf C})] /\Omega_N(z), \eeq
where $\Omega_N(z)$ is the grand partition function for a ring of $N$ sites.

The grand partition function $\Omega_N (z)$ can be determined using the 
standard transfer matrix technique. We find the largest eigenvalue of the 
$2 \times 2$ matrix ${\bf T}_2$ given by
\beq {\bf T}_2 = \left[ \begin{array}{cc}
1 & 1 \\ z & 0 \\ \end{array} \right]. \eeq
We now calculate $\langle \psi | H_d | \psi \rangle$. The matrix element of 
the $i$-th term is clearly zero, unless $n_{i-1} = n_{i+1} =0$. Then the only
non-zero matrix element is 
\beq \langle H_d \rangle / N = - 2 \sqrt{z} ~{\rm Prob}(000) = 
\frac{-2}{\sqrt{z}} {\rm Prob}(010). \eeq
Here ${\rm Prob}(000)$ denotes the probability that randomly selected three 
consecutive sites in the ring will be empty in the classical ensemble, and 
similar definition for ${\rm Prob}(010)$. This is easily calculated for the 
Hamiltonian $H_{cl}$ in the limit of large $N$. We get
\beq {\rm Prob}(010) = {\rm Prob}(1) = \rho. \eeq
The largest eigenvalue $\Lambda$ of ${\bf T}_2$ is given by
\beq \Lambda = ( 1 + \sqrt{ 1 + 4 z})/2, \eeq
and $\rho$ is the density per site given by $\rho = z d\log (\Lambda)/ dz$.
Extremizing $\langle H_d \rangle$ with respect to $z$, we find that the 
minimizing value occurs for $z= 0.405$, yielding $\langle H_d \rangle = 
-0.60057$. This gives us the variational bound 
variational bound on the ground state energy per site $E_0$ 
\beq E_0 \leq -0.60057. \label{eq39} \eeq
This energy is somewhat higher than the energy obtained in the previous 
section (see Fig. \ref{eigenvaluesfit}), indicating
that the correlations in the classical Hamiltonian $H_{cl}$ do not exactly 
reproduce the correlations in the quantum ground state of $H_d$.

We can make a better variational calculation by considering a classical 
lattice gas with an additional next-nearest-neighbor interaction. The 
Hamiltonian of this lattice gas is
\beq H'_{cl} = + \infty \sum_i n_i n_{i+1} - K \sum_i n_i n_{i+2} - \mu 
\sum_i n_i. \label{hpcl} \eeq
Let us denote $z = \exp(\beta \mu)$, and $u = \exp( \beta K)$. In this case,
the transfer matrix is a $3 \times 3$ matrix given by
\beq {\bf T}_3 = \left[ \begin{array}{ccc}
1 & 0 & 1 \\ z & 0 & z u \\ 0 & 1 & 0 \\ \end{array} \right]. \eeq
The probability of the configuration ${\bf C}$ in the 
equilibrium ensemble is given by
\beq {\rm Prob}({\bf C}) = \exp[ -\beta H'_{cl}({\bf C})] / \Omega_N(z,u), \eeq
where $\Omega_N(z,u)$ is the grand partition function for a ring of $N$ sites.
We then get
\bea - \langle H_d \rangle / N &=& 2 {\rm Prob}(00000) \sqrt{z} + 4 {\rm Prob}
(10000) \sqrt{zu} \non \\
& &+ 2 {\rm Prob}(10001) \sqrt{z u^2}, \eea
Here ${\rm Prob}(00000)$ is the probability of finding a randomly selected 
set of five consecutive sites all unoccupied in the equilibrium ensemble 
corresponding to the Hamiltonian $H'_{cl}$. These probabilities are also easily 
calculated. Treating $z$ and $u$ as variational parameters, we find that 
$\langle H_d \rangle$ is minimized for $z = 0.35198$ and $u = 1.3752$. For 
these values one finds that the density is $\rho = 0.1952$, ${\rm Prob}(00000)
=0.28066$, ${\rm Prob}(10000)=0.082804$, and ${\rm Prob}(10001) = 0.02443$. 
These give
\beq E_0 \leq -0.60333 \eeq
which is an improvement over Eq. (\ref{eq39}), and quite close to the 
extrapolated value of $-0.60356$ obtained from Table I. This may be further 
improved by taking third-neighbor interactions in the classical Hamiltonian, 
but this will not attempted here.

\section{Study of ground states in other sectors}
\label{dilute}

We define a more general Hamiltonian
\beq H(\lambda) = H_{Kit} + \lambda \sum_n W_n. \eeq
Since the $W_n$'s commute with $H_{Kit}$, all the eigenvectors of $H_{Kit}$ 
can be chosen to be simultaneous eigenvectors of $H(\lambda)$, for all 
$\lambda$. However, if we vary $\lambda$, we can get different eigenvectors 
to have the lowest energy. 

Clearly, if $\lambda$ is large and positive, the ground state will lie in the 
sector with all $W_n =-1$. Conversely, if $\lambda$ is large and negative, 
the ground state is the lowest energy eigenvector in the sector with all $W_n 
= +1$. In both these regions, the gap in the excitation spectrum is of order 
$|\lambda|$. As we vary $\lambda$ from $-\infty$ to $+\infty$, initially the 
gap decreases and becomes zero at some value $\lambda^c_1$. We then 
expect a gap to open up again when $\lambda$ is greater than a second 
critical point $\lambda^c_2 > \lambda^c_1$.

\subsection{Sectors with most $W_n$'s positive}

Since the ground state for $\lambda =0$ lies in the sector with all $W_n =+1$,
we have $\lambda^c_1 > 0$. In fact, if the lowest excitation energy in the
Hamiltonian $H_{Kit}$ is $\Delta E$, we have $\lambda^c_1= \Delta E/2$. At
this point, the energy required to change a single $W_n$ from $+1$ to $-1$ 
becomes zero. We now study this sector using the variational techniques of 
Sec. V and try to estimate the difference between the ground state energy of 
this sector and the sector with all $W_n =1$.

Without loss of generality, we may assume that in this sector, $W_N= -1$, and
the rest of the $W$'s are +1. The basis vectors in this sector are of type
$|xU\rangle$, or $|Vy\rangle$, where $U$ and $V$ are all possible strings of
length $N-1$ obtainable from the string $zzz \ldots z$ of length $N-1$, using
the substitution rule $zz \to yx$. Let $|\psi\rangle$ be the
eigenvector corresponding to the lowest eigenvalue of $H$ in this
sector. It is easy to verify that $\langle xU|H|xU'\rangle$ and $\langle
Vy|H|V'y\rangle$ are negative, for all $U$ and $U'$, and $V$ and $V'$. 
But, $\langle Vy|H|xU\rangle$ are positive. This implies that $\langle
xU|\psi\rangle$ and $\langle xU'|\psi\rangle$ have the same sign for all $U$
and $U'$. Similarly $\langle Vy|\psi\rangle$ and $\langle V'y|\psi\rangle$ have
the same sign for all $V$ and $V'$. This suggests a variational wave function
of the form 
\beq |\psi \rangle =\frac{1}{\sqrt{2}} \left[ \sum_U \sqrt{{\rm Prob}(U)} |xU
\rangle - \sum_V \sqrt{ {\rm Prob}(V)} |Vy\rangle \right]. \eeq

Here ${\rm Prob}(U)$ and ${\rm Prob}(V)$ are arbitrary functions, satisfying 
the constraint
\beq \sum_U {\rm Prob}(U) = \sum_V {\rm Prob}(V) =1. \eeq

Each configuration $U$ is in one-to-one correspondence with the configurations
of a nearest-neighbor-exclusion lattice gas on a linear chain of length $(N-
2)$. Define a chain configuration as $C= \{n_i\}$. We put $n_i =1$ if and only
if there is a $y$ in $U$ in the position $i+1$, otherwise $n_i=0$. Note that 
the last element of $U$ cannot be a $y$. We specify $C$ by a binary string of 
length $(N-2)$. To go from $C$ to $U$, we first add a single $0$ to the binary
string of $C$ at the right end, and then use the substitution rule $10 \to
yx$. The remaining zeros in $C$ are replaced by $z$'s. Similarly, we specify
$V$ also by a binary string of length $(N-2)$, with $x \to 1$, $~y, z
\to 0$, and as the leftmost element of the resulting string is always
a zero, it may be deleted.

As in the previous calculation, we construct a classical Hamiltonian to
variationally estimate the parameters ${\rm Prob}(C)$. In this case, there 
is no translational symmetry, and in general, the lattice gas will have a 
non-trivial density profile. This is taken into account by making the
activities of the lattice gas in the classical Hamiltonian site-dependent. 
We write 
\beq H_{cl}^C = + \infty \sum_{i=1}^{N-3} n_i n_{i+1} - \sum_{i=1}^{N-2} 
\mu_i n_i. \eeq
The probability of each configuration $C$ of the lattice gas is then given by
\beq \textmd{Prob(C)}= \exp[-\beta H_{cl}^C (C)]/\Omega_{N-2}(\{ z_i\}), \eeq
with $z_i = \exp( \beta \mu_i)$, and $\Omega_{N-2}(\{z_i\})$ is the grand 
partition function of the open chain of $N-2$ sites. 

We note that the matrix $H$ is unchanged under the space reflection
$i \leftrightarrow N+1 -i$, and at the same time exchanging $x$ and $y$.
This can be built into our eigenvector by assuming that if $V$ are strings 
corresponding to lattice gas configurations $C$, we set 
\beq {\rm Prob} (V) = {\rm Prob} (U), \eeq
where $U$ is the string corresponding the lattice gas configuration $C^T$, the
transpose of $C$. 

The rest of the calculation is done as before. By construction, we have 
\beq \langle \psi_{\textmd{var}}| \psi_{\textmd{var}} \rangle = 1. \eeq
It is straightforward to express $\langle \psi| H| \psi \rangle$ in terms of
the marginal probabilities of the different local configurations of the lattice
gas, remembering that there is no translational invariance. For example, we get
\beq \langle \psi| S^x_N S^y_1 | \psi \rangle = - {\rm Prob}_C(n_1 = 0). \eeq

In the simplest case, we work with only two parameters, and set $z_1 = z'$,
and $z_i = z$ for $i \neq 1$. We would like to estimate the difference of the
ground state energy in this sector and the ground state over all sectors. These
energies are of order $N$, and to cancel the leading linear $N$-dependence, we
have to set $z$ equal to the optimal value $z^* = 0.4045$ to get the best
energy value of Eq. (\ref{eq39}). We assume that $N$ is large, so that only
the term in the partition function corresponding to the largest eigenvalue is 
kept. Extremizing over $z'$ we obtain $z'= 0.2537$, and for this value
\beq \langle \psi_{\textmd{var}}|H|\psi_{\textmd{var}} \rangle = -0.6005 N + 
0.1875. \eeq 
This implies the following bound on the lowest eigenvalue in this sector
\beq E'_0 \leq N E_0 + \Delta. \eeq with $\Delta = 0.1875$ providing an 
estimate of the energy gap between the ground state and 
the first excited state of the Hamiltonian.
This estimate can be improved by adding more parameters in the variational
wave function, or equivalently in the classical lattice gas Hamiltonian. A two
parameter wave function would have $z'$ and $z''$ at the two opposite ends.
Extremizing with respect to these parameters we find the energy gap to be
$0.1642$ with $z' = 0.2537$ and $z'' = 0.6670$. A four parameter wave function
would have fugacities $z_1,z_2, z_{N-3}$ and $z_{N-2}$ adjustable, and the rest
of the $z_i$'s set equal to $z^*$. Table \ref{delta} shows the improvement in 
the value of the energy gap with the number of parameters used.

We thus obtain a variational estimate of the energy gap of the first excited 
state from the ground state energy. This matches quite well with the numerical
estimates obtained in the previous section.

\begin{table}[h]
\begin{center} \begin{tabular}{|c|c|} \hline
Number of & $\Delta$ \\ 
parameters & \\ \hline
1 & 0.18751 \\ \hline
2 & 0.16419 \\ \hline
4 & 0.15845 \\ \hline
6 & 0.15642 \\ \hline
8 & 0.15578 \\ \hline
10 & 0.15556 \\ \hline
\end{tabular} \end{center}
\caption{Estimate of the energy gap with the number of parameters used.} 
\label{delta} \end{table}

It is straightforward to extend this treatment to sectors with two or more 
$W_n$'s negative. There is an energy $\Delta$ required to create a single 
negative $W_n$. Thus $\lambda^c_1 = \Delta/2$. If two defects are spaced far 
apart, the energy required to 
create two defects will be nearly $2 \Delta$, with the correction term 
decreasing exponentially with the distance between the defects. For $n$ 
defects, the energy would be minimized if the defects are equally spaced. 
Thus the distance between the defects is $N/n$, and the energy cost of 
creating $n$ defects $\Delta E(n)$ in $H(\lambda)$, for small $n$, is well 
approximated by 
\beq \Delta E(n) ~\approx ~- ~2 n \lambda +~ n \Delta +~ n A \exp (-BN/n), 
\label{expbn} \eeq
where $A$ and $B$ are some constants. This then implies that for $\lambda = 
\lambda^c_1 + \epsilon$, the density of defects in the true ground state of 
$H(\lambda)$ will vary as $1/|\log \epsilon|$.

\subsection{Sectors with most $W_n$'s negative}

We now discuss the behavior of the ground state energy near the critical 
point $\lambda^c_2$. This depends on the behavior of the ground state energy 
in sectors in which only a few of the $W_n$'s are $+1$. 

For a ring of $N$ sites, the sector with all $W_n = -1$ contains only two 
states, $xxxxx \cdots$ and $yyyyyy \cdots$. The two are degenerate, with 
eigenvalue equal to $- \lambda N$.

Now consider the sector with only one $W_n = +1$, say $W_0 =+1$. Consider 
the state $\psi_1 =|zxxxx \cdots \rangle$ in this sector. From Eqs. 
(\ref{sxsy2}), under $H_{Kit}$, we have $zx \rightleftharpoons yz$, and this 
state can make a transition only to the state $\psi_2 = 
|yzxxx \cdots \rangle$. And $\psi_2$ can return to $\psi_1$ or go to $\psi_3 
= |yyzxx \cdots \rangle$. Thus, the dynamics may be considered as the 
dynamics of a particle $z$, which can hop to a nearest neighbor under the 
action of the Hamiltonian. There is a string of $y$'s connecting the current 
position of the particle to the leftmost allowed position which is $n=1$. 
This string can become longer, or shorter, as the particle moves, with no 
energy cost. When the $z$-spin is at the site $N$, it cannot move further to 
the right. The ground state energy $E^g_{1-sector}$ if this sector is seen to 
be the same as that of a particle with nearest-neighbor hopping, confined to 
move in the space $1 \leq x \leq N$. It is thus given by
\beq E^g_{1-sector} = - (N-2) \lambda - 2 J \cos (\frac{\pi}{N+1}). \eeq
Thus, we see that for large $N$, the state with all $W_n$'s equal to +1 is 
no longer the ground state for $\lambda < J$.

We now consider a sector with exactly two of the $W_n$'s equal to $+1$, 
and the rest negative. Let us start with the state $|zxxx \cdots zxxx \cdots 
\rangle$, where the spins at two sites $i=1$ and $i=m+1 
\leq N$ are in the state $z$ (these states will be referred to as 
$z$-spins in the following). This corresponds to $W_N = W_m =+1$. Then, 
under the action of $H(\lambda)$, this state mixes with other states 
where the positions of the $z$-spins can change; the general state in 
this sector may be labeled by the positions of the $z$-spins, $r_1$ and 
$r_2$. We will write the vector as $|r_1,r_2\rangle$, where $1\leq r_1 
\leq m < r_2 \leq N$. Then, for $1 < r_1 <m$ and $m+1 < r_2 < N$, we get
\bea H_{Kit} |r_1,r_2\rangle &=& -|r_1, r_2+1\rangle -|r_1+1,r_2\rangle \non \\
& & - |r_1, r_2-1\rangle - |r_1-1, r_2\rangle. \eea
 
If the first $z$-spin is at $m$ and the second is not at $m+1$, the first spin
cannot move to $m+1$, as that site would be in spin state $y$, and the state 
$zy$ cannot change [Eqs. (\ref{sxsy1})]. Similarly, if $r_2= N$, and $r_1 \neq
1$, then the second spin cannot move to the right. However, if the two 
$z$-spins are adjacent, then they can change to a state $zz \rightleftharpoons
yx$ [Eqs. (\ref{sxsy1})]. But from the state $yx$ the state can only return to
$zz$.

If we disallow the transitions to state $yx$, the $z$-spins act as independent 
particles moving in two disjoint regions of space, $1 \leq r_1 \leq m$ and 
$m+1 \leq r_2 \leq N$. In this case, the minimum energy of this system is 
just the sum of the energies of two particles. This 
energy is an upper bound on the true ground state energy of this system. 
Thus, we find that the ground state energy in this sector, $E^g_{2-sector}$, 
has the upper bound
\bea E^g_{2-sector} &\leq& -2 J \cos(\frac{\pi }{ m+1}) - 2 J \cos 
(\frac{\pi}{N-m+1}) \non \\
& & - \lambda (N-4). \eea

Next, suppose that the state with the $m$-th site in the $y$-state and the
$(m+1)$-th in the $x$-state is called the state $r_1 = m+1, r_2 =1$, and a 
similar definition for the other end. Then the range of $r_1$ is at most 
$m+1$, and the range of $r_2$ is at most $l-m+1$. By excluding some states 
(here $r_1 =m+1, r_2 \neq r_1$), the kinetic energy can only increase, and 
hence we have
\bea E^g_{2-sector} &\geq& -2 J \cos(\frac{\pi }{ m+2}) - 2 J \cos 
(\frac{\pi}{N-m+2}) \non \\ 
& &- \lambda (N-4). \eea
For $N, m \gg 1$, these bounds can be expanded in powers of $1/m$, and have 
the same leading order correction. Also, the minimum energy corresponds to 
equally spaced defects, with $m=N/2$.

We can easily extend the discussion to sectors with three, four or more 
$W_n$'s equal to $+1$. In case the lengths of the intervals between the 
positive $W_n$'s are $m_1, m_2, m_3, \cdots, m_r$, the bounds on the lowest 
energy in this sector $E^g_{r-sector}$ become
\bea -2 J \sum_{i=1}^r \cos( \frac{\pi}{m_i+2}) - \lambda(N-2r) \leq 
E^g_{r-sector} \non \\
\leq -2 J \sum_{i=1}^r \cos( \frac{\pi}{m_i+1}) -\lambda(N-2r). \eea

Thus, we see that for $\lambda > J$, the ground state belongs to the 
sector with all $W_n$'s equal to $-1$. If $\lambda = J(1 - \ep)$, the ground 
state will be in the sector with $n$ equispaced bonds with $W_n =+1$, where 
the spacing $\ell$ between them $\approx N/n$ is given by $\ep^{-1/2}$. 
The minimum energy per site of $H(\lambda)$ for $\lambda = J(1 -\ep)$ 
varies as $\ep^{3/2}$ for small $\ep$. Equivalently, if we restrict 
ourselves to sectors with only a fraction $\ep$ of $W_n$'s having the value 
$+1$, the minimum energy per site varies as $- \ep^{3/2}$. This is equivalent 
to the statement that for $H_{Kit}$ corresponding to $\lambda =0$, in the 
sector with the fractional number of positive $W_n$'s being equal to 
$\Delta$, the minimum energy per site varies as $\Delta^{3/2}$. 

\section{Discussion}
\label{disc}

In this paper, we first analyzed the symmetries of a spin-$S$ Kitaev chain. 
We found a $\mathbb{Z}_2$ invariant, $W_n$, associated with every link 
$(n,n+1)$, namely, $N$ invariants for the model defined on a ring with $N$ 
sites. For integer $S$, these
invariants commute with each other and the Hamiltonian. The Hilbert space can
therefore be split into $2^N$ sectors, where the Hamiltonian is block diagonal.
For half-odd integer $S$, $W_n$ anti-commutes with $W_{n\pm 1}$ and commutes
with the rest. We showed that this implies that all the eigenstates of the
half-odd-integer spin models are $2^{N/2}$-fold degenerate, thus showing a
qualitative difference between the integer and half-odd-integer models. We have
developed a formalism to compute the dimensions of the invariant sectors. We 
showed that the dimension of most of the sectors can be calculated in terms of 
products of $2 \times 2$ matrices ${\mathbb T}_+$ and ${\mathbb T}_-$. For 
$S=1$ the quantum dimension of the sector with all $W_n=1$ is the golden 
ratio, $(1+\sqrt{5})/2$. For $S\to\infty$, the quantum dimension tends to 
$S+1/2$ in both the $W_n=1$ and the $W_n=-1$ sectors. 

We have then studied the spin-1 case in detail. We have found that the
ground state lies in a sector which can be mapped to a quantum lattice
gas model with nearest-neighbor exclusion. We developed a variational
wave function that relates the quantum mechanical averages to the
correlation functions of a classical lattice gas with nearest-neighbor
exclusion. We considered a more general Hamiltonian with a term
proportional to the sum of the conserved quantities, and showed that as a
function of the coupling constant $\lambda$, this would show gapless
excitations in the range $\lambda^c_1 \leq \lambda \leq \lambda^c_2$.
We extended our variational calculation to study how the ground state
energy and the defect density would vary near the two critical points
$\lambda^c_1$ and $\lambda^c_2$. At $\lambda= \lambda^c_1$, Eq. (\ref{expbn}) 
implies that the energy of the lowest excited state in a system of length $L$ 
goes as $E \sim \exp (-BL)$, corresponding to a state in which one $W_n = -1$ 
while all the other $W_n = 1$. By the usual scaling arguments, the gap to the 
first excited state goes as $1/L^z$, where $z$ is the dynamical critical 
exponent. We therefore conclude that $z = \infty$. At $\lambda = \lambda^c_2$,
the low-energy excitations form a low-density 
gas of hard-core particles. In one dimension, this can be mapped to a system 
of non-interacting spinless fermions with a non-relativistic spectrum $E \sim 
k^2$. Hence in a system of size $L$, the gap to the lowest energy states
goes as $1/L^2$ corresponding to $k \sim 1/L$; thus $z = 2$. It would be 
interesting to find the value of $z$ in the critical region $\lambda^c_1 
< \lambda < \lambda^c_2$.


Finally, we note that there is another interesting one-dimensional spin model 
called the golden or Fibonacci chain \cite{feiguin,trebst}, for which the 
number of states on a ring of size $N$ is the same as that of the spin-1 Kitaev
chain in the sector with all $W_n = 1$. The Hamiltonian for this model is
\bea \non
H_{GC}&=& \sum_i ~\left( (n_{i+1}+n_{i-1}-1) \right. \\ \label{gchain}
&& \left. -n_{i-1}n_{i+1} (\gamma^{-3/2}\sigma^x_i+\gamma^{-3}n_i+1+
\gamma^2) \right), \non \\
&& \eea
where $n_i=(1-\sigma^z)/2$. It has been shown \cite{feiguin,trebst} that this
model is critical. Its long-range correlations are described by a $SU(2)$ level
3 Wess-Zumino-Witten model, which is a conformally invariant field theory with
central charge equal to $7/10$. The Hamiltonian in Eq. (\ref{gchain}) differs 
from the spin-1 Kitaev chain in the $W_n=1$ sector by terms which are products
of the $n_i$ operators. We have shown that the spin-1 Kitaev chain is gapped. 
Thus these terms correspond to some relevant operators which take the golden 
chain Hamiltonian away from criticality.

We can show that it is possible to add multi-spin
terms to the minimal Kitaev chain which reduce to the extra terms in the
$W_n=1$ sector. We need to add products of the $n_i$ 
operators to the minimal Kitaev chain to obtain the golden chain in the sector
with all $W_n= 1$. The basis states $\vert\uparrow\rangle$, $\vert\downarrow
\rangle$ that we use in Sec. \ref{shalf} are eigenstates of the $n_i$ operators
with eigenvalues $1$ and $0$ respectively. The $\vert\uparrow\rangle$ state 
represents a state with the head, namely $\vert x\rangle$. The 
$\vert\downarrow\rangle$ state represents either an empty site, 
$\vert z\rangle$, or a tail, $\vert y\rangle$. It is clear from Eqs. 
(\ref{sigx1}) and (\ref{xyzketdef}) that the operator 
$P^x\equiv(1+\Sigma^x)/2$ has eigenvalues $1$ for $\vert x\rangle$ and $0$
for $\vert y\rangle$ and $\vert z\rangle$. Since all the $\Sigma$ matrices 
commute for integer $S$, they commute with the invariants and are block
diagonal within the invariant sectors. Thus, the Hamiltonian,
\bea \non
H_{KGC}&=&\gamma^{-3/2}H_{KC}-\sum_i\left(1-P^x_{i+1}-P^x_{i-1}\right.\\
\label{kcgc} 
&&\left. +\gamma^2P^x_{i-1}P^x_{i+1} +\gamma^{-3}P^x_{i-1}P^x_i P^x_{i+1}
\right), \eea
when restricted to the $W_n=1$ sector, is exactly the golden chain Hamiltonian
discussed by Feiguin {\it et al.} and others \cite{feiguin,trebst}. We have 
thus constructed a realization of the golden chain model as a spin-1 chain.

\section*{Acknowledgments}

We thank G. Baskaran for interesting comments. DS thanks DST, India for 
financial support under Project No. SR/S2/CMP-27/2006. DD thanks DST, India 
for support through a J. C. Bose Fellowship under SR/S2/JCB-24/2006.

\end{document}